\documentclass[sigconf]{acmart}
\AtBeginDocument{%
  }

\copyrightyear{2026}
\acmYear{2026}
\setcopyright{cc}
\setcctype{by}
\acmConference[ReCode '26]{1st Workshop on Code Translation, Transformation, and Modernization }{April 12--18, 2026}{Rio de Janeiro, Brazil}
\acmBooktitle{1st Workshop on Code Translation, Transformation, and Modernization (ReCode '26), April 12--18, 2026, Rio de Janeiro, Brazil}
\acmPrice{}
\acmDOI{10.1145/3786180.3788315}
\acmISBN{979-8-4007-2411-4/2026/04}

\begin{document}

\title[Environment-in-the-Loop]{Environment-in-the-Loop: \\ Rethinking Code Migration with LLM-based Agents}

\author{Xiang Li}
\affiliation{%
  \institution{University College London}
  \city{London}
  \country{United Kingdom}}
\email{x.li.25@ucl.ac.uk}

\author{Zhiwei Fei}
\affiliation{%
  \institution{Nanjing University}
  \city{Nanjing}
  \country{China}}
\email{zhiweifei@smail.nju.edu.cn}

\author{Ying Ma}
\affiliation{%
  \institution{Brunel University}
  \state{London}
  \country{United Kingdom}}
\email{ying.ma@brunel.ac.uk}

\author{Jerry Zhang}
\affiliation{%
  \institution{Delysium}
  \city{Panama}
  \country{Republic of Panama}}
\email{support@delysium.com}

\author{Sarro Federica}
\affiliation{%
  \institution{University College London}
  \city{London}
  \country{United Kingdom}}
\email{f.sarro@ucl.ac.uk}

\author{He Ye}
\affiliation{%
  \institution{University College London}
  \city{London}
  \country{United Kingdom}}
\email{he.ye@ucl.ac.uk}
\renewcommand{\shortauthors}{Xiang Li et al.}

\begin{abstract}
Modern software systems continuously undergo code upgrades to enhance functionality, security, and performance, and Large Language Models (LLMs) have demonstrated remarkable capabilities in code migration tasks. However, while research on automated code migration which including refactoring, API adaptation, and dependency updates has advanced rapidly, the exploration of the automated environment interaction that must accompany it remains relatively scarce. In practice, code and its environment are intricately intertwined. Relying solely on static analysis of the environment leads to an inadequate understanding of the target setting, prolongs feedback cycles, and consequently causes significant rework and project delays, thereby reducing overall efficiency. We contend that successful software evolution demands a holistic perspective that integrates both code and environment migration. To understand the current landscape and challenges, we first provide an overview of the status of automated environment construction. We then propose a novel framework paradigm that tightly integrates automated environment setup with the code migration workflow. Finally, we explore the challenges and future directions for automated environment interaction within the code migration domain. Our findings emphasize that without automated environment interaction, the automation of code migration is only half complete.
\end{abstract}


\maketitle

\section{Introduction}

Code migration is a critical and frequent challenge in modern software development, which involves large-scale and systematic modifications across an entire codebase at the repository level particularly. Code migration tasks represent a comprehensive evolution of software across its language, version, system, and dependencies. Typical scenarios include: 1) API and dependency upgrades; 2) framework modernization (e.g. from AngularJS to React); 3) language version evolution (e.g. from Python 2 to Python 3, or from Java 8 to Java 17); 4) platform and system porting (e.g. from Ubuntu to CentOS). The high degree of difficulty in these tasks stems from their broad impact and strong inter-dependencies. Developers must maintain consistency across numerous files, understand complex cross-module call relationships, and handle the cascading effects on external dependencies, database schemas, and deployment scripts. Consequently, establishing a stable and reliable post-migration environment is the cornerstone of a successful migration.

\begin{figure}[!t]
  \centering
  \includegraphics[width=0.8\linewidth]{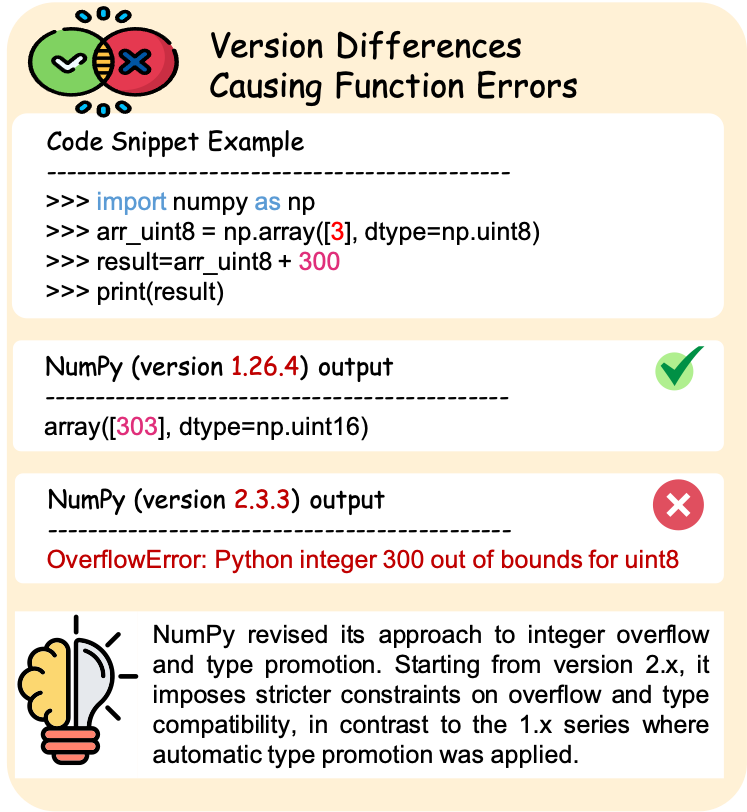}
  \caption{An implicit and subtle runtime error caused by the library upgrade, which is hard to catch by static analysis.}
  \vspace{-1em}
  \label{fig:example}
\end{figure}

To manage the complex engineering of code migration, the process is typically structured into several key phases~\cite{cheng2025codemenvbenchmarkinglargelanguage,fahmideh2020challengesmigratinglegacysoftware,LinearB2025CodeMigration,may2025freshbrewbenchmarkevaluatingai,10.1145/3613905.3650896}. The current basic steps for code migration include: 1) examine the existing system’s structure and environment to define migration scope and plan; 2) update code to resolve incompatibilities and replace deprecated callings; 3) execute and test the migrated system to ensure functional correctness.

\begin{figure}
  \centering
  \includegraphics[width=0.9\linewidth]{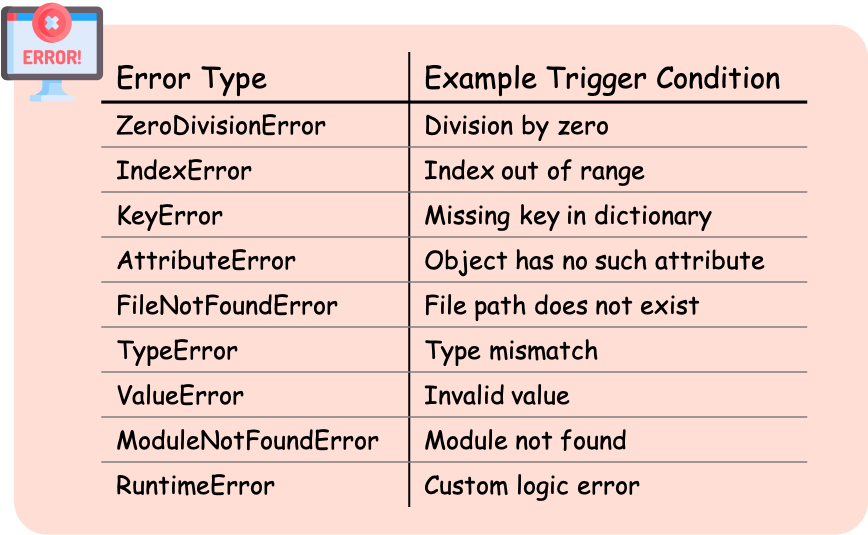}
  \caption{Examples of error types that can only be discovered through dynamic environment interaction.}
  \vspace{-1em}
  \label{fig:errors}
\end{figure}

Current migration processes overly focus on code modification, and rely on static analysis of the initial environment, decoupling code modification from environment configuration. This disconnect misses subtle runtime errors caused by different library versions, which static analysis cannot detect. As shown in Fig.\ref{fig:example}, although the function name and usage remain the same, different library versions (version 1.x and 2.x of NumPy) may impose different internal constraints, resulting in subtle errors that cannot be caught by static analysis alone. Fig.\ref{fig:errors} shows some examples of hidden and subtle errors that cannot be identified by static environment analysis timely. Furthermore, this approach mistakenly treats the environment as immutable, ignoring that code logic itself often dictates environmental changes (like library upgrades or downgrades). This leads to unnecessary rework and project delays ~\cite{Macwan2025, Jan2025ConfigDrift, PuppetConfigDrift2023}.

Although Large Language Models (LLMs) demonstrate exceptional capabilities in software engineering tasks~\cite{chen2025prometheus}, and are used in migration ~\cite{Jelkhoury2025LLMCodeMigration, LinearB2025CodeMigration, 10.1145/3597503.3639226}, their application is confined to interactive code editing. As reported by ~\citet{cheng2025codemenvbenchmarkinglargelanguage}, LLMs perform poorly in predicting execution outcomes (causing nearly 30\% of runtime errors), indicating a need for actual environment interaction. Research on how to leverage LLMs to interact with environments still needs to be explored. The key challenge is to research how LLMs can automate environment setup, code execution and testing, thereby establishing a dynamic and closed-loop feedback between the code and its environment across the migration process.

To address this, we propose an \textbf{LLM-based environment-driven migration framework} to replace the current linear and manual process.  Rather than depending on initial static analysis, it creates a dynamic feedback loop and feeds real-time, practical information from the environment (e.g., test failures, runtime behavior) back into the migration process to guide and refine the code modification. This integrated approach allows for the early detection and resolution of problems, significantly reducing rework while increasing the overall efficiency and success rate of the migration.

\section{Background}\label{sec:background}

\subsection{Challenges}
The core deficiency in current code migration workflows stems from the disconnection between the code modification process and the dynamic realities of the execution environment. The primary challenges can be summarized as follows:
\begin{itemize}
    \item Decoupling from the execution environment. The over-reliance on static analysis of the initial environment, detached from the actual migration steps, makes it impossible to detect subtle, version-dependent runtime errors (e.g., internal constraint changes between NumPy 1.x and 2.x) that are imperceptible to static checks. This runtime error can lead to a large amount of rework~\cite{thiyagarajan2024ai,fahmideh2020challengesmigratinglegacysoftware}. For example, Running the test/CI/build phases after AI translates/migrates code may expose unexpected errors/dependency incompatibility issues, which often lead to the need to go back and modify and update the translated code~\cite{Ziftci_2025}. 
    \item Manual validation and test workflow. Post-migration validation, including environment setup, dependency resolution, and testing, remains a labor-intensive, manual process~\cite{cheng2025codemenvbenchmarkinglargelanguage, yuan2025semanticalignmentenhancedcodetranslation}. This workflow erroneously treats the environment as immutable, ignoring that code logic changes often necessitate corresponding environmental adaptations (e.g., library upgrades), thereby causing significant rework~\cite{Ziftci_2025,wang2025codesyncsynchronizinglargelanguage, horton2019v2fastdetectionconfiguration, kontogiannis2010code}.
    \item Absence of a dynamic feedback loop. The most critical research gap is the lack of a mechanism to leverage LLMs for automated environment interaction~\cite{10.1145/3613905.3650896, chen2025fortran2cppautomatingfortrantoctranslation}. This includes automated setup, code execution, and results-in-formed testing. 
\end{itemize}

\subsection{Automated Environment Interaction}

Creating a functioning development environment for a codebase or a repository, is a vital task in software development. Rule-based or template-based automated environment construction and interaction ~\cite{bndr_pipreqs, horton2019dockerizemeautomaticinferenceenvironment, microsoft_generator‐docker} can enhance efficiency and provide valuable assistance, but usually require a significant amount of manual input from developers or are limited to specific languages or use cases. Recently, the extensive research interest in AI agents has promoted its application in software engineering, especially in the field of automated environments interaction. LLM-based agents typically consist of four key components: planning, memory, perception, and action ~\cite{xi2023risepotentiallargelanguage}, which ensure its efficient and accurate automation process in the field of environmental interaction. ~\cite{milliken2025beyond,hu2025llm} propose LLM-based agent capable of successfully setting up Python repositories. ~\cite{bouzenia2025you} introduce ExecutionAgent that correctly configures 33 out of 50 considered repositories across 5 programming languages (Python, Java, C, C++, and JavaScript). EnvBench ~\cite{eliseeva2025envbench} is a comprehensive environment setup benchmark, which encompasses thousands of Python and JVM-based repositories, with a focus on repositories that present genuine configuration challenges. Nowadays, LLM-based agents have demonstrated superior performance compared to standalone LLMs in various software engineering tasks~\cite{xu2025aligningobjectivellmbasedprogram, chen2025prometheus}, but few researches have discussed how agents affect migration fields.

\begin{figure*}[!t]
  \centering
  \includegraphics[width=0.9\linewidth]{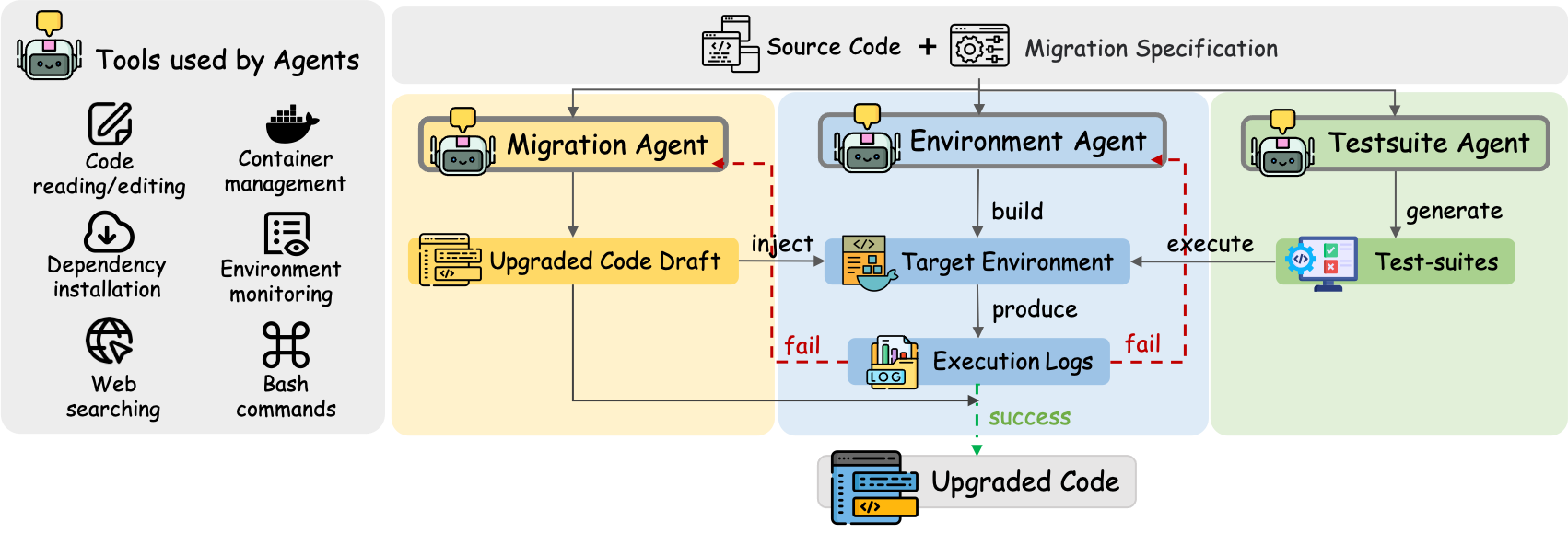}
  \caption{The overall workflow of environment-driven multi-agent code migration.}
  \label{fig:main_workflow}
  \vspace{-1em}
\end{figure*}

\subsection{LLM-assisted Code Migration}

The early work ~\cite{horton2019v2fastdetectionconfiguration} point out that re-configurability of the environment is a core prerequisite for migration automation. LLMs has brought about a new intelligent path for code migration.~\citet{Jelkhoury2025LLMCodeMigration} and ~\citet{LinearB2025CodeMigration} demonstrate that LLMS can understand complex project structure and dependency contexts, leading to fewer warnings, and lower risks. But LLMS only act as "advisors" rather than "executors" in these cases, and the interactive potential has yet to be fully exploited.~\citet{cheng2025codemenvbenchmarkinglargelanguage} build a benchmark to verify LLMs' abilities in code upgrades, downgrades, and version migrations, in which the result indicates that LLMs have initial ability to understand environmental constraints and perform functional verification. However, the target environment itself is often unknown or dynamically generated in real project migration.

In industrial practice, an increasing number of migration tools have moved beyond merely providing static code-change suggestions. For instance, Amazon Q’s ~\cite{10.1145/3613905.3650896} \textit{dev mode} and Aviator Agent’s Runbooks framework ~\cite{AviatorDocs} can launch an isolated sandbox (Docker container) where candidate changes are built and tested; if tests fail, the system iterates improvements. When integrated into GitHub, CI/workflow results feed back into the Agent’s suggestion process. These industrial examples jointly support a clear emerging trend: future migration tools should bridge code, environment, and runtime validation into a unified loop, thereby improving the accuracy, automation capability, and reliability of migrations.

Altogether, existing research has not yet established a systematic paradigm for an Agent that actively builds environments and interacts with them to carry out code migration. So we want to propose such a new paradigm that enabling an Agent not only to understand code, but also to manipulate the environment (install, retry, correct, test), thereby realizing a truly end-to-end automated migration system.

\vspace{-1em}
\section{Environment-driven Multi-Agent Workflow}\label{sec:workflow}
The migration workflow operates as a multi-agent collaborative loop, centered on continuous iteration across four stages: 1) migration planing, 2)automated environment setup, 3)test validation, 4)feedback refinement, as demonstrated in Fig.\ref{fig:main_workflow}. Each agent undertakes a distinct responsibility within the migration chain, while the chain synchronizes their actions to achieve end-to-end migration verification.

\vspace{-1em}
\subsection{Agent Components}
We abstract the LLM-based Migration process into an Agent system composed of three core collaborative components: Migration Agent (M-Agent), Environment Agent (E-Agent), Testsuite Agent (T-Agent). Among them, each agent can use a specific functional toolkit (see left part of Fig.\ref{fig:main_workflow}), responsible for handling specific stage tasks in the migration workflow.

\textbf{Migration Agent.} The Migration Agent (M-Agent) is responsible for understanding the migration goal and program semantics, and for performing code modification and migration. Its inputs include the source project code and the migration target specifications (e.g., language version, framework upgrade, API compatibility changes). Its outputs are candidate migrated code versions, dependency upgrade plans, and lists of potential risk points. Within the workflow, the M-Agent recieve feedback from the E-Agent to automatically refine the migrated code logic and dependency mapping strategies.

\textbf{Environment Agent.} The Environment Agent (E-Agent) autonomously constructs executable build and runtime environments for the migrated project, ensuring that the transformed code can be correctly compiled, executed, and validated under real dependency configurations. It takes as input the migrated repository, dependency manifests, and runtime constraints (such as language version, platform, and operating system). Its outputs include successfully built environment images, runtime logs, diagnostic reports for environment-related errors, test execution results and failure reports. In the collaborative workflow, the E-Agent acts as the central verification hub: 
\begin{itemize}
\item It provides the execution context required by the T-Agent for testing.
\item It supplies environmental feedback and error diagnostics.
\item It triggers remediation loops in either the M-Agent (for code-level errors) or itself (for configuration-level issues).
\end{itemize}

By enabling reproducible and traceable build environments, the E-Agent ensures migration reliability across dependency and configuration drifts, transforming the process from static code transformation into a dynamic, environment-aware migration cycle.

\textbf{Testsuite Agent.} The Testsuite Agent (T-Agent) is responsible for automatically generating, repairing, or extending test suites to validate functional consistency and behavioral equivalence after migration. Its inputs include the migrated codebase (from the M-Agent), existing legacy tests, behavioral specifications, and project documentation. Its outputs are generated test files (unit, integration, and regression tests). Through iterative execution within the E-Agent’s constructed environment, the T-Agent continuously improves validation coverage and provides structured behavioral feedback to guide further refinement.

\subsection{Migration Workflow}
In this workflow, the Environment Agent is not merely a supporting component but the central driver of automated migration. By grounding the entire migration life-cycle in automated environment setup, the proposed paradigm advances code migration from static transformation toward operational intelligence.

\subsubsection{Task Trigger and Migration Planning}
The workflow is automatically triggered when version upgrades, framework migrations, or dependency changes are detected within the CI/CD pipeline. The M-Agent is firstly invoked to analyze the repository and synthesize an initial migration plan, producing a migrated code draft and updated dependency specifications. These outputs are then immediately handed to the E-Agent, which becomes the central executor for all subsequent validation and feedback.

\subsubsection{Automated Environment Interaction}
Upon receiving the migrated repository and dependency manifests, it autonomously constructs a reproducible, executable environment—including language runtimes, dependency packages, toolchains, and system-level configurations. Concretely, the E-Agent:
1) selects base images and provisioning scripts, 2) installs dependencies and resolves version conflicts, 3) configures build systems (e.g., Gradle/Maven, npm/pip), and 4) builds and executes the project in an isolated sandbox.

If the build or runtime fails, the E-Agent captures logs, diagnoses causes (e.g., compiler mismatch, missing packages, dependency incompatibility), and generates structured diagnostic reports. These reports form environmental feedback, which drives the entire corrective process: 
\begin{itemize}
    \item Configuration-level errors are self-repaired by the E-Agent.
    \item Semantic or logic-level issues are routed back to the M-Agent.
    \item Runtime behavioral anomalies are passed to the T-Agent for validation refinement.
\end{itemize}

Through this process, the Environment Agent transforms the migration from a static, one-time code transformation into a dynamic and verifiable process, where every migration step is grounded in an executable, environment-aware state.

\subsubsection{Test Generation and Execution}
Once a stable environment is established, the T-Agent automatically generates regression test-suites by analyzing the migrated code and legacy behavioral specifications. All tests are executed within the E-Agent’s verified environment, ensuring that test outcomes reflect realistic dependency and configuration constraints. Crucially, the E-Agent captures runtime logs, records system states, and provides contextual traces in the execution loop for each failure.

\subsubsection{Feedback Aggregation and CI/CD Integration}
After each test iteration, all results are analyzed with the environmental context as the anchor point. This iterative cycle is orchestrated around the E-Agent, whose diagnostics determine the next correction target. The E-Agent thereby evolves into a continuous verification and coordination hub, maintaining the consistency of the migration pipeline across agents.

Upon successful validation, the migrated code version, environment configurations, generated tests, coverage reports and execution logs will be aggregated and uploaded to CI/CD artifact repository. This enables automatic revalidation not only during version upgrades but also whenever dependency or configuration drift is detected.

The entire closed-loop workflow which includes three LLM-based agents achieves end-to-end automation from migration analysis to environment construction, testing, and iterative repair. This paradigm ensures migration correctness while enhancing CI/CD resilience against environment drift and compatibility degradation.
\vspace{-1em}
\section{Challenges and Further Directions}\label{sec:challenges}

Despite the promising potential of the proposed agentic migration paradigm, several open challenges remain, especially regarding automation and reliability of environment setup. Environment reproduction and dependency management remain core difficulties. Reconstructing a runnable environment that faithfully mirrors the original system demands accurate detection of configuration drift and robust dependency inference. Future work may enable Environment Agents to learn environment semantics from build scripts, runtime traces, or CI logs to ensure reproducible, portable environments. Multi-agent coordination and feedback loops pose another challenge. Environment feedback (e.g., build errors, missing libraries) must be interpreted by Migration Agents to trigger meaningful fixes rather than simple retrials. Future research could draw on reinforcement learning or cooperative multi-agent systems to build more stable correction cycles.

Industrial integration with CI/CD pipelines is also crucial. Automated environment construction must align with existing workflows, supporting sandbox creation, reconstruction logging, and developer auditing. Future directions include leveraging LLM-based Environment Agents as transparent extensions to container orchestration systems such as Docker or Kubernetes.

\vspace{-1em}
\section{Conclusion}\label{sec:conclusion}
We take the position that automated environment setup must become a first-class concern in the design of code migration systems. Current research and industrial tools focus predominantly on source code, yet overlook the fact that migration correctness is ultimately verified only within a runnable environment. We advocate a paradigm shift: migration should be as a co-evolution of source code, dependencies, and execution environments. By embedding this capability into CI/CD workflows and coupling it with continuous feedback from Migration and Testsuite Agents, we envision a future where code migration becomes an environment-intelligent process, which capable of sustaining long-term software evolution across ecosystems.

\bibliographystyle{ACM-Reference-Format}
\bibliography{sample-base}

\appendix

\end{document}